\documentstyle[prd,aps,floats]{revtex}

\begin{document}
\draft

\input epsf \renewcommand{\topfraction}{0.8}
\twocolumn[\hsize\textwidth\columnwidth\hsize\csname
@twocolumnfalse\endcsname

\title{Scalar field potentials for cosmology}
\author{V\'{\i}ctor H. C\'{a}rdenas and Sergio del Campo}
\address{Instituto de F\'{\i}sica, Pontificia Universidad Cat\'{o}lica de Valpara\'{\i}so (PUCV), Casilla 4059,\\
Valpara\'{\i}so, Chile}
\date{\today}
\maketitle

\begin{abstract}
We discuss different aspects of modern cosmology through a scalar
field potential construction method. We discuss the case of
negative potential cosmologies and its relation with oscillatory
cosmic evolution, models with a explicit interaction between dark
energy and dark matter which address the coincidence problem and
also the case of non-zero curvature space.
\end{abstract}

\pacs{PACS numbers: 98.80.Cq \hfill GACG-04-01 } \vskip2pc]


\section{Introduction}

In the last five years there have been increasing observational
evidence that our universe seems to be dominated by a unknown
component with a negative pressure usually called dark energy or
quintessence. This component is the source for the accelerated
expansion that Type Ia supernovae measurements implies. Taken
together both the CMB experiments and the Type Ia supernovae
results, indicates that our universe is almost flat $\Omega
_{0}=1.03\pm 0.05$ with a matter contribution of $\Omega
_{m}=0.29$ and a cosmological constant component with $\Omega
_{\Lambda }=0.73$ \cite{data}. The problem with assuming the
existence of a true non-zero cosmological constant is the fact
that theoretical considerations estimate its contribution in more
than $100$ orders of magnitude the measurable value. There are two
main alternatives to deal with this: the first one, try to explain
it as a modification of Einstein's theory of gravitation \cite
{Modif} and the other is to assume that the true cosmological
constant is zero, and work with the idea that this small
contribution is given by unclumped energy in the form of a scalar
field which has not reached its ground state \cite{ini}. The first
alternative is still under early research and nothing crucial has
been found yet. The second proposal has been studied by several
years finding interesting results, however this picture has two
main problems: first of all, the mass of the field has to be
extremely small to keep the field rolling to its vacua even today,
and second the model require we live in a special time: it is just
the time when dark energy start to dominate the energy density of
the universe, which is known as the coincidence problem.

Almost all the attempts to address this problem require a scalar field
usually called quintessence or dark energy with the special property to
trace the matter energy density evolution \cite{ini},\cite{qtsnc}, leading
to an asymptotic constant ratio $\Omega _{m}/\Omega _{\Lambda }$. This has
been done by using special scalar potentials like $V=V_{0}\exp (1/\varphi )$
or $V=V_{0}\varphi ^{-\alpha }$ where the parameter $V_{0}$ has to be
adjusted to the critical energy today. This implies of course that we are
not solved the problem, we have just shifted it. However some of these
tracker solutions \cite{qes}, \cite{inter}, those where the scalar field
potential is an exponential function of $\varphi $, the energy density
always follows the matter energy density evolution. The problem with this
kind of tracking field is the possibility to potentially modify the Big Bang
Nucleosynthesis (BBN) calculations and also the problem of having a scalar
field today tracking matter implying a wrong equation of state.

In a recent paper Padmanabhan \cite{padma} presented a simple set
of equations to construct a scalar field potential $V(\varphi )$
given a particular form of evolution for the scale factor $a(t)$.
The analysis was done in a flat universe filled with both a scalar
field $\varphi $ alone and also with an extra matter component
present. Even earlier, Ellis and Madsen \cite{ellis} have
discussed such a reconstruction program with an special emphasis
in closed universes. Although observational evidence seems to
confirm the inflationary prediction of a spatially flat universe,
the data marginally prefer a closed geometry. This fact has led
some to consider the consequences of a non zero curvature in the
present universe. In this context recently there have been
interest in inflationary and quintessence models with arbitrary
curvature \cite{closed} and also to consider the consequences of
an explicit interaction between baryonic/dark matter and dark
energy, which also has been considered in the literature
\cite{inter}. This lead us to consider a generalization of the
procedure or ``recipe'' taking into account interaction between
matter and the dark energy and also a non-zero curvature.

In this paper we perform a reconstruction scheme for the scalar
field potential in a non zero curvature universe with a explicit
interaction term between matter components. The basic elements of
the method are presented in section II while section III is
dedicated to negative potentials. We find that asymptotic
exponential potentials are usually the generic solution but
depending on what boundary conditions we impose is possible to get
a broad class of potentials. The interaction between matter and
dark energy components is studied in section IV and the case of
non-zero curvature is developed in section V. We find that
although a general recipe is always possible to construct, in the
general case is not easy to find explicit solutions. We summarize
our conclusions in section VI.

\section{The basic recipe}

To keep things clear, in this section we concentrate in the case
without interaction. Assuming a FRW metric with arbitrary
curvature the Friedman equation and the scalar field equation are
\begin{equation}
\left( \frac{\dot{a}}{a}\right) ^{2}+\frac{k}{a^{2}}=\kappa ^{2}\rho ,
\label{eq1}
\end{equation}
\begin{equation}
\ddot{\varphi}+3H\dot{\varphi}+V^{\prime }(\varphi )=0,  \label{eq2}
\end{equation}
where $H=\dot{a}/a$ is the Hubble parameter, $\kappa ^{2}=8\pi G/3$ and $%
\rho =\dot{\varphi}^{2}/2+V(\varphi )$ is the scalar field energy density.
By multiplying Eq. (\ref{eq2}) by $\dot{\varphi}$ we can rewrite it in its
fluid form as
\begin{equation}
\dot{\rho}+3H(\rho +p)=0,  \label{eq3}
\end{equation}
where the density pressure is $p=\dot{\varphi}^{2}/2-V(\varphi )$. By using
Eq.(\ref{eq1}) we can rewrite Eq.(\ref{eq3}) to find
\begin{equation}
\frac{\dot{\rho}}{\rho }=2H\left[ \frac{\dot{H}a^{2}-k}{(Ha)^{2}+k}\right] .
\label{eq4}
\end{equation}
Assuming for the scalar field an equation of state of the form $p=\omega
(\varphi )\rho $ we can combine Eq.(\ref{eq3}) and (\ref{eq4}) to get
\begin{equation}
1+\omega (\varphi )=-\frac{2}{3}\left[ \frac{\dot{H}a^{2}-k}{(Ha)^{2}+k}%
\right] .  \label{eq5}
\end{equation}
From the definitions of the energy density and pressure of the scalar field
we can write $\dot{\varphi}^{2}/2V(\varphi )=(1+\omega )/(1-\omega )\equiv
f(t)$ and from this definition and using \ Eq.(\ref{eq2}) is easy to show
that
\[
\frac{\dot{V}}{V}=-\frac{\dot{f}+6Hf}{1+f}.
\]
From this equation is possible to write down directly the relations between
the scalar field potential $V(\varphi )$ and the field $\varphi $ in terms
of $t$ as
\begin{equation}
V(t)=\frac{3}{8\pi G}\left( H^{2}+\frac{\dot{H}}{3}+\frac{2k}{3a^{2}}%
\right) ,  \label{eq6}
\end{equation}
and
\begin{equation}
\dot{\varphi}^{2}(t)=-\frac{1}{4\pi G}\left( \dot{H}-\frac{k}{a^{2}}\right) .
\label{eq7}
\end{equation}
We can check that these equations are equal to the equations
derived in \cite {padma} in the case $k=0$, and also coincides
with those showed in \cite {ellis} and \cite{Ellis2}. If we add an
extra known component $\rho _{m}(t)$ with density pressure $p_{m}$
we can modify directly these equations as follows. From here on an
explicit subscript is written to differentiate both components.
The Friedman equations is modified by
\[
H^{2}+\frac{k}{a^{2}}=\frac{8\pi G}{3}\left( \rho _{m}+\rho _{\varphi
}\right) =\frac{8\pi G}{3}\rho _{\varphi }(1+r)
\]
where $r(t)=\rho _{m}/\rho _{\varphi }.$ From this arrangement Eq.(\ref{eq4}%
) gets an extra term $\dot{r}/3H(1+r)$ in its right hand side and finally
\begin{equation}
V(t)=\frac{3(1+r)^{-1}}{8\pi G}\left( H^{2}+\frac{\dot{H}}{3}+\frac{2k%
}{3a^{2}}-\frac{\dot{r}}{6(1+r)}\right) ,  \label{eq8}
\end{equation}
and the scalar field
\begin{equation}
\dot{\varphi}^{2}(t)=\frac{(1+r)^{-1}}{4\pi G}\left( -\dot{H}-\frac{k}{a^{2}}%
+\frac{\dot{r}H}{2(1+r)}\right) .  \label{eq9}
\end{equation}
We can also check that this is the generalization of the relations found in
Ref.\cite{padma} for both arbitrary curvature and an extra matter component.
In that paper a new function $Q(t)$ is defined associated with the extra
matter component. In our notation we prefer to parametrize the new matter
component defining the function $r(t)=\Omega_{m}/\Omega_{\varphi}$ which is
more appropriated to address the coincidence problem. Of course both are
simply related through $r(t)=3H^{2}Q(t)/8\pi G\rho_{\varphi}$.

\section{The case of negative potentials}

As a first example, let us consider the case of a oscillating universe whose
evolution is given by $a(t)=a_{0}\sin (\alpha t)$. We have a flat geometry $%
k=0$ with just a scalar field as a matter content. Although naive,
this kind of evolution has been considered several times in the
past and more recently in the cyclic or ekpyrotic model
\cite{cyclic}. If we assume that nothing else than the scalar
field is present, as most of the studies on negative potentials
do, we can use the construction equations (\ref{eq8}) and (\ref
{eq9}) and find
\begin{equation}
\int d\varphi (t)=\left( \frac{\alpha ^{2}}{4\pi G}\right)
^{1/2}\int dt\left( 1+\cot ^{2}\alpha t\right) ^{1/2},
\end{equation}
assuming that we cancel out the constant factor due to the integration, we
find that
\begin{equation}
\varphi (t)=-\left( \frac{1}{4\pi G}\right) ^{1/2}\cot (\alpha t),
\end{equation}
leading to a potential of the form
\begin{equation}
V(\varphi )=\frac{\alpha ^{2}}{4\pi G}(\sinh ^{2}(\sqrt{4\pi G}\varphi )-1),
\end{equation}
which near the origin evolves as the worked example considered in
Ref.\cite{negp}, i.e., $V(\varphi )=m^{2}\varphi ^{2}/2+V_{0}$
with $V_{0}<0$, where
\begin{equation}
m^{2}=\alpha ^{2}\text{, and}\hspace{0.5cm}V_{0}=-\alpha ^{2}/4\pi G
\end{equation}
This potential is particularly interesting because the evolution
of the scale factor is very simple. An interesting thing with this
potential model is the possibility to study its dynamical
evolution and critical points to determine its attractors and
fixed points. This work is under study \cite {VCJS}. There is also
the possibility to have a non-singular evolution of the scale
factor like $a(t)=A+B\sin(\alpha t)$, where $A$, $B$ and $\alpha$
are constants which keeps $a(t)>0$. In this case we can use the
construction equations to determine the field
\begin{equation}
\sqrt{4\pi G}(\phi - \phi_o)= B\alpha \int
\frac{\sqrt{1+(A/B)\sin(\alpha t)}}{A+B\sin(\alpha t)}dt
\end{equation}
In general this integral leads to elliptic functions, and is
difficult to extract more information about the potential.
However, we can study numerically its behavior to extract the
shape of the potential. For a sample of parameter the potential
looks like Fig.1. Again the scalar field potential take negative
values near its minimum.
\begin{figure}[tb]
\centering \leavevmode\epsfysize=6cm \epsfbox{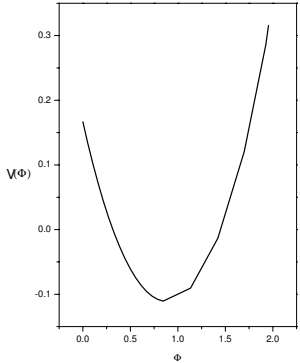}
\caption{This graph shows the integrated scalar field potential
obtained for a non-singular universe in a flat universe. Notice
that again the potential has negative values near the origin.}
\end{figure}
We notice that it seems possible to generate a cousin model, for
every model based on a closed geometry, by using negative
potentials. In this context, we can also study a model similar to
the ``Emergent Universe'' proposed in \cite{Ellis3}. In that
reference is assumed that our universe is closed and the authors
worked out an example where the universe start in a static
Einstein state which evolve towards an expanding phase that leads
to inflation. Based on our results in this section it seems
probable to construct a similar model in a flat universe but with
negative potentials.

\section{Quintessence-matter interaction}

In the previous section we derived a set of equations for a simple non flat
model where a scalar field and matter have an asymptotic constant energy
densities ratio $r(t)\equiv \Omega _{m}/\Omega _{\varphi }$. \ This model
can be easily generalized to the case where a explicit interaction is added
between matter and dark energy.

If we consider an explicit interaction, then Eq.(\ref{eq2}) and the equation
of motion of the fluid become modified by
\begin{equation}
\ddot{\varphi}+3H\dot{\varphi}+V^{\prime }(\varphi )=\zeta ,  \label{eq10}
\end{equation}
\begin{equation}
\dot{\rho}_{m}+3H(\rho _{m}+p_{m})=-\zeta \dot{\varphi},  \label{eq11}
\end{equation}
where $\zeta $ specify the interaction, and $\rho_{m}$ and $p_{m}$
are the matter energy and pressure densities. There are many ways
to couple these components, for example in \cite{qes} Wetterich
uses $\zeta =\alpha \rho _{m} $ resembling the coupling between
baryons and nucleons in GUT theories and Chimento et
al.\cite{inter} use $\zeta =3\Pi H/\dot{\varphi}$ where $\Pi $ is
certain density dissipative pressure. Another class of interaction
is of the type $\zeta =-\Gamma \dot{\varphi}$ \ where $\Gamma $ is
a constant which depends on physical parameters of the particle
model. This kind of interaction is known in the theory of
reheating where $\Gamma $ is interpreted as the rate of particle
decay from the inflaton field to radiation. For simplicity, in the
rest of the section we restrict the analysis to a universe with
$k=0$.

Let us start the analysis with $\zeta =\alpha \rho _{m}$. The
Friedman equation is the same as Eq.(\ref{eq1}) with $k=0$ where
now $\rho =\rho _{\varphi }+\rho _{m}$. Essentially the key
expression to obtain here is the analog of Eq.(\ref{eq4}).

From the Friedman equation we obtain
\begin{equation}
\frac{\dot{\rho}_{\varphi }}{\rho _{\varphi }}=2\frac{\dot{H}}{H}-\frac{\dot{%
r}}{(1+r)}  \label{eq12}
\end{equation}
From Eq.(\ref{eq10}) we can calculate the same ratio as
\begin{equation}
\frac{\dot{\rho}_{\varphi }}{\rho _{\varphi }}=-3H(1+\omega (\varphi
))+\alpha \dot{\varphi}r  \label{eq13}
\end{equation}
Because we are interested in the asymptotic evolution of the
system, and we want to obtain a scalar field potential with the
property of tracking behavior, we set $r=r_{0}=$\- constant for
the rest of the analysis. This assumption enable us to obtain a
universe without the coincidence problem. To extract more
information from this method we have to assume also the evolution
of the matter density. Actually this is what must be done if we
want to perform the same analysis in section II with a variable
$r(t)$. This is not so tricky because we are looking for
asymptotic solutions where we assume the scalar field and the
matter density have already reached smooth evolutions. If this
assumption is not correct, then can not get consistent solutions.
So, let us consider the case where $\rho _{m}\sim t^{-q}$. It
immediately implies a temporal evolution for the time derivative
of the
scalar field, because we have assumed before that $r=r_{0}$, so $\dot{\varphi%
}\sim t^{-q/2}$. This also affects the evolution of
Eq.(\ref{eq11}). From this we obtain that $\alpha
\dot{\varphi}\sim t^{-1}$, implying $q=2$. Of course we want to
obtain a solution of the form similar to Eq.(\ref{eq9}), so the
previous expression for $\dot{\varphi}$\ has to be taken as an
ansatz. The final result has to be consistent with this analysis.

Combining Eq.(\ref{eq12}) and (\ref{eq13}) we obtain the generalization of
Eq.(\ref{eq4})
\begin{equation}
1+\omega (\varphi )=\frac{1}{3H}\left[ -\frac{2}{3}\frac{\dot{H}}{H}+\frac{%
r_{0}C}{t}\right] .  \label{eq14}
\end{equation}
where $C=\alpha \dot{\varphi}_{0}$. From this relation and by inspection of
Eq.(\ref{eq5}) is direct to obtain the case where $k\neq 0$. The formulaes
for constructing the potential and scalar field are in this case
\begin{equation}
V(t )=\frac{3H^{2}}{8\pi G}\left( 1+\frac{1}{3}\frac{\dot{H}}{H^{2}}-%
\frac{r_{0}C}{6Ht}\right) ,  \label{eq15}
\end{equation}
and
\begin{equation}
\dot{\varphi}^{2}(t)=\frac{3H^{2}}{8\pi G}\left( -\frac{1}{3}\frac{\dot{H}}{%
H^{2}}+\frac{r_{0}C}{6Ht}\right) .  \label{eq16}
\end{equation}
If we consider a power law evolution, $a(t)=a_{0}t^{n}$, then we find that $%
H=nt^{-1}$ and $\dot{H}=-nt^{-2}$, so replacing in Eq.(\ref{eq15}) and (\ref
{eq16}) we find
\[
V(\varphi )=V_{0}\exp (-\lambda \varphi ),
\]
with
\begin{eqnarray*}
V_{0} &=&\frac{1}{8\pi G}\left( (3n-1)n+\frac{nr_{0}C}{2}\right) , \\
\lambda  &=&\sqrt{8\pi G}\left( n(1+\frac{r_{0}C}{2})\right) ^{-1/2},
\end{eqnarray*}
and the scalar field evolve as
\[
\varphi (t)=\sqrt{\frac{n}{8\pi G}\left( 1+\frac{r_{0}C}{2}\right) }\ln t,
\]
consistent with our ansatz. We may think that this method is not so helpful
because from the very beginning we have found $\dot{\varphi}\sim t^{-1}$ and
also, from the definition of $\rho _{\varphi }$, the relation $V\sim t^{-2}$%
, leading to an exponential potential. But this is just an ansatz
and we need the complete solution to understand all the
consequences.

\section{Non-zero curvature case}

In this section we perform the application of the equations derived in
section I. Let us assume first a universe without interaction between matter
components, $\zeta =0$. As a first example, we study the well known problem
of a closed ($k=1$) oscillatory universe, that with a scale factor evolving
as $a(t)=a_{0}\sin (\alpha t)$. If a matter component is present we assume
that the ratio $r(t)=r_{0}$ is a constant. By introducing the scale factor
in the reconstruction equations we find
\begin{eqnarray}
V(\varphi ) &=&\frac{3(1+r_{0})^{-1}}{8\pi G}\times  \label{pot} \\
&&\times \left[ \frac{2}{a_{0}^{2}}-\frac{\alpha ^{2}}{3}+2\left( \frac{1}{%
a_{0}^{2}}+\frac{\alpha ^{2}}{3}\right) \sinh (\left| \varphi /\varphi
_{c}\right| ),\right]  \nonumber
\end{eqnarray}
where we have define
\[
\varphi _{c}=\sqrt{\frac{1}{4\pi G}\left( \frac{\alpha ^{2}-a_{0}^{-2}}{%
1+r_{0}}\right) .}
\]
The scalar field derivative is
\[
\dot{\varphi}^{2}(t)=\varphi _{c}^{2}(1+\cot ^{2}(\alpha t)).
\]
A subtle treatment of this equation leads to the $\left| \varphi \right| $\
dependence of $V(\varphi )$. After we take the square root, we can define
two branches: those where $\dot{\varphi}>0$, and $\dot{\varphi}<0$. Taking
care of this conditions we found that
\[
\sinh \alpha \left| \frac{\varphi }{\varphi _{c}}\right| =\cot (\alpha t).
\]
The potential (\ref{pot}) looks like Fig. 2.
\begin{figure}[tb]
\centering \leavevmode\epsfysize=6cm \epsfbox{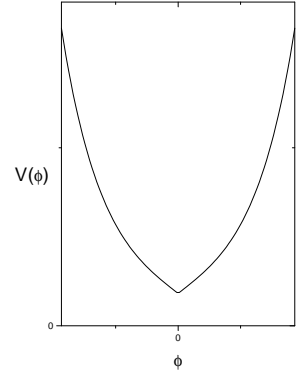}
\caption{This graph shows a typical scalar field potential used in
a closed inflationary model. The scale is arbitrary.} \label{fig1}
\end{figure}
 Of course, this potential can also drive chaotic inflation. The
Emergent Universe \cite{Ellis3} is a special case of a system
where we impose a solution like $a(t)=A+B\exp(\alpha t)$.

\section{Results}

The method presented in this paper looks extremely important
theoretically for the study of model building in cosmology.
Although it is not the first paper on this subject we have
generalized the recipe to construct scalar field potential for
arbitrary curvature and also with an explicit interaction between
matter and the scalar field.

The interaction between matter components give us the possibility
to address the coincidence problem without appealing to any other
tracker property. In fact, in the case of tachyons, the tracker
evolution does not work but is possible to get a similar tracking
behavior by using an explicit interaction.

By considering non-zero curvature lead us to many different models
where is possible to integrate the evolution. In particular,
during the last years there have been some controversy about
considering inflation in a closed geometry \cite{closed}. In this
paper we shows a explicit model where inflation can be obtained
followed by a oscillating reheating phase. The Emergent Universe
proposed in \cite{Ellis3} is a special case of a system where we
impose a solution like $a(t)=A+B\exp(\alpha t)$.

The applications are multiple; inflationary building models and
also the more recently dark energy problem or quintessence.
Although is always possible to use the construction equations
found in this paper, we stress that in general we can not extract
a explicit shape of the potential. Probably the main interest in
this kind of construction technic is the possibility to extract
some information about the shape of the potential in certain
ranges and then look for those features in well motivated
theoretical models.

\section{Acknowledgments}

The authors would like to thank Ioav Waga for useful conversations
and comments. This work has been supported by projects FONDECYT
grant 3010017 (VHC) and 1030469 (SdC). Also, it was partially
supported by UCV grant No. 123.764/03.

\end{document}